\documentclass[fleqn,twoside]{article}
\usepackage[dvips]{graphicx}
\usepackage[usenames,dvips]{color}
\usepackage{url}
\usepackage{cite}
\usepackage{amsmath}
\usepackage{rotating}
\usepackage{espcrc2}

\newcommand{\red}[1]{{\color{Gray} #1}}

\def\be#1\ee{\begin{equation}#1\end{equation}}
\def\bea#1\eea{\begin{eqnarray}#1\end{eqnarray}}
\def\ba#1\ea{\begin{align}#1\end{align}}
\def\btab#1\etab{\begin{table}[th]\begin{center}#1\end{center}\end{table}}
\def\bfig#1\efig{\begin{figure*}[th]\begin{center}#1\end{center}\end{figure*}}
\def\bi#1\ei{\begin{itemize}#1\end{itemize}}
\def\bc#1\ec{\begin{center}#1\end{center}}

\def\unix{{\sc Unix}}

\def\diana{{\sc Diana}}
\def\form{{\sc Form}}
\def\fortran{{\sc Fortran}}

\def\ff{{\sc FF}}
\def\LT{{\sc LoopTools}}
\def\FA{{\sc FeynArts}}
\def\FC{{\sc FormCalc}}
\def\qgraf{{\sc Qgraf}}
\def\hahn{\texttt{Hahn}}
\def\aitalc{{\sc \textit{a}{\r{\i}}\raisebox{-0.14em}{T}alc}}

\title{
Fully automated calculation in fermion scattering
\thanks{Work supported in part by European's 5-th Framework under
  contract HPRN--CT--2000--00149 Physics at Colliders and by the
  Deutsche Forschungsgemeinschaft under contract SFB/TR 9--03.}
}

\author{A. Lorca\address[zeuthen]{Deutsches Elektronen-Synchrotron,
    DESY, Platanenallee 6, 15738 Zeuthen, Germany}
}

\begin{document}

\begin{abstract}
The package \aitalc\ has been developed for fully automated calculations of
two fermion production at $e^+ e^-$ collider and other similar reactions. We
emphasize the connection and interoperability between the different modules
required for the calculation and the external tools \diana, \form\ and
\LT. Results for $e^+ e^- \rightarrow f \bar{f}, e^+ e^-$ are presented.
\end{abstract}

\maketitle

\section{\label{introduction}INTRODUCTION}
The proposed $e^+ e^-$ international linear collider (ILC), is
expected to open a new era of precision physics and help understanding
many of the fundamental questions that the discoveries of the
forthcoming large hadron collider (LHC) will arise. With energies up
to at least 500 GeV in its first phase \cite{ILCTRC}, the ILC will
uniquely provide a level of experimental precision not reached so far
yet in particle physics, demanding an equivalent understanding of the theoretical
predictions and uncertainties which, for the two fermion production, should not be larger than a few permill.

In order to fulfill such theoretical duty, the calculations
within the present standard model must include
typically the next or next-to-next to leading order perturbative
corrections, depending on the process. The case of two fermion
production is specially interesting since it plays a key role for 
\begin{enumerate}
\item luminosity monitoring (Bhabha scattering),
\item directly measuring precise distributions and asymmetries,
\item establishing limits on {\em New Physics} as contact interaction, $Z'$, leptoquarks, or gravitons in extra dimensions,
\item quantifying backgrounds for other processes.
\end{enumerate}

Thanks to the recent progress in computer technologies and software
applications, particle physics has greatly profited from an
indeterminate number of tools and packages developed for the specific
task of automating perturbative calculations. Globally speaking,
today's available calculational tools do not individually
meet the next properties:
\begin{itemize}
\item Performing higher order corrections beyond leading order.
\item Free-availability and free-based code (i.e.\ neither the package
  nor its software requirements have to be purchased).
\item Comprehensibility, both in documentation and installation.
\item Flexibility, being able to deal with different processes, theoretical models and to allow some modifications.
\end{itemize}

Next, it will be shortly described some features of the package
\aitalc\, showing some interesting immediate applications in two fermion production.

\section{\label{automation}AUTOMATION}
The tool \aitalc\ \cite{Lorca:2004fg} is designed with the intention
of matching the desired requirements pointed out above. Its goal is to perform
automated perturbative calculations of cross sections in high energy
physics, restricted to $2 \rightarrow 2$ fermion processes within the
electroweak standard model (EWSM) and quantum electrodynamics (QED).

The tool integrates only free-of-charge packages existing on the
market. As the calculation proceeds, in a modular fashion, it profits
from an adaptation \cite{Fleischer:2004tg} of the \diana\ package \cite{Tentyukov:1999is} (based itself on the \qgraf\ \cite{qgraf} code) for the generation and analysis of Feynman graphs, the \form\ \cite{Vermaseren:2000nd} language when dealing analytically with the large expressions in the amplitudes and, finally, the \LT\ \cite{Hahn:1998yk} library (also integrating the \ff\ package \cite{vanOldenborgh:1990yc}) for the numerical calculation of the loop integrals.

In order to simplify and guarantee a wide portability during
installation, it proceeds conforming to the standards of the \unix\
packages. The operations include a customization of your system
settings via {\tt configure} and the later extraction of individual
packages (if needed) and examples by a {\tt make} instruction. At this
point the installation is finished.

A detailed discussion of the algorithms used, the strategy of the
calculation and internal work of each of the modules has been already presented
elsewhere \cite{Lorca:2004fg,Lorca:2004dk,Lorca:2005yp}. Here I will shortly
discuss the working environment and the examples provided to the user.

\subsection{Makefile as environmental language}
Automation requires effective intercommunication and a smart way to
organize the tasks. This second concept has been exploited through the
Makefile environment, providing a chain dependency between files and
modules. The Makefile environment accomplishes the running of the
whole process, from letting the user only the responsability of
modifying the driver file to building the {\tt tree}, {\tt loop} and
{\tt fortran} sections. At the end, the compilation of the program
leading to the final results is achieved without further interaction.

The main directory into a process contains a main Makefile that
organize the calculation in a generic footing. Under a generic unique
{\tt make} command, it calls the submakefiles {\tt Level.mk} and {\tt
  Fortran.mk}, providing them the necessary variables to carry out the
tree level, the loop level and the compilation of the generated code
for numerical output. The Makefile will consider the inclusion of
extra files in the main process directory as new settings for the
computation instead of the default ones appearing in the directories {\tt form/kitform3/} and {\tt diana/prg/}.

As a result of the previous call the directories {\tt tree/}, {\tt
  loop} and {\tt fortran} are created and filled with the log
files of the diagram generation, symbolic amplitudes, eps drawings
of the diagrams (for the two levels) and the complete \fortran\
programs, subroutines with the form factors grouped by topologies and
other kinematical functions. The elapsed time will always be printed
in the file {\tt Makefile.log}.

The numerical executable program {\tt main.out} is located inside {\tt
fortran} and on its default running returns a sample of differential
cross sections for five different angles at a given energy of 500
GeV. There are 7 printout columns: first and second remind the energy
and scattering angle. Born cross section, virtual loop
corrections and soft photon radiation appear in the columns 3 to 5
respectively, while the sum of all them corresponds to the column number 6
(the expected result). The last column is reserved for the
squared of the loop amplitude serving as a rough estimator of the order of the
error at next perturbative order, or main result in cases where the tree
level is not physically possible\footnote{As in the flavour changing neutral
currents processes}. Alternative integrated cross sections and/or
forward-backward asymmetries are available when the logical
variables {\tt lprintics}, {\tt lprintfba} become `true'. For them a
second group of columns show the estimated integration error following
Richardson extrapolation to Romberg integration \cite{Richardson}.

\subsection{Examples with loop corrections}
The package contains three examples located in the {\tt
  examples/} directory which the user may run without further
considerations for test and learning purposes. Two of them contain radiative corrections:
\begin{itemize}
\item Bhabha scattering in QED ($e^- e^+ \rightarrow e^- e^+$ in {\tt Bhabha\_QED}),
\item flavour changing through neutral current example ($e^- e^+
  \rightarrow b \bar{s}$ in {\tt leLe.bS}).
\end{itemize}

With the purpose of giving a flavour of execution time, the examples
were run on two different machines, representing today's typically
available computers. The results are shown in Table {timings} and take
from some seconds to minutes. Only when the size of the produced
\fortran\ code begins to be large (i.e.\ more than $\sim$200KB for a
single subroutine to be compiled), then the GNU compiler gets
extremely overloaded slowing down significantly the overall time.

\btab
\caption{Typical timings for the different modules. The technical specifications for system `Desk(top)': Intel Pentium III 853MHz cpu, 256MB RAM, GNU {\tt g77} version 3.3.3 compiler. For `Lap(top)' are: Intel Centrino 1.5GHz cpu, 512MB RAM, Intel {\tt ifort} version 8.1 compiler.}
\label{timings}
\begin{tabular}{lrrrr}
\hline
&\multicolumn{2}{c}{\tt leLe.bS}&\multicolumn{2}{c}{\tt bhabha\_QED}\\
Module&Desk&Lap&Desk&Lap\\
\hline
{\tt tree}&4''&1''&30''&9''\\
{\tt loop}&2'20''&52''&37''&12''\\
{\tt fortran}&4:35'56''  &33''&26''&8''\\
\hline
Total&4:38'20''&1'26''&1'33''&29''\\
\hline
\end{tabular}
\etab

\section{Selected results}

\subsection{Fermion anti-fermion production}
The $e^-e^+ \rightarrow b\bar{b}$ and $e^-e^+ \rightarrow
t\bar{t}$ are two of the processes where it is expected to measure,
with relatively high statistics, the indirect impact of the Higgs
sector due to the large couplings involved through the top quark mass.

\bfig
\includegraphics[scale=0.9]{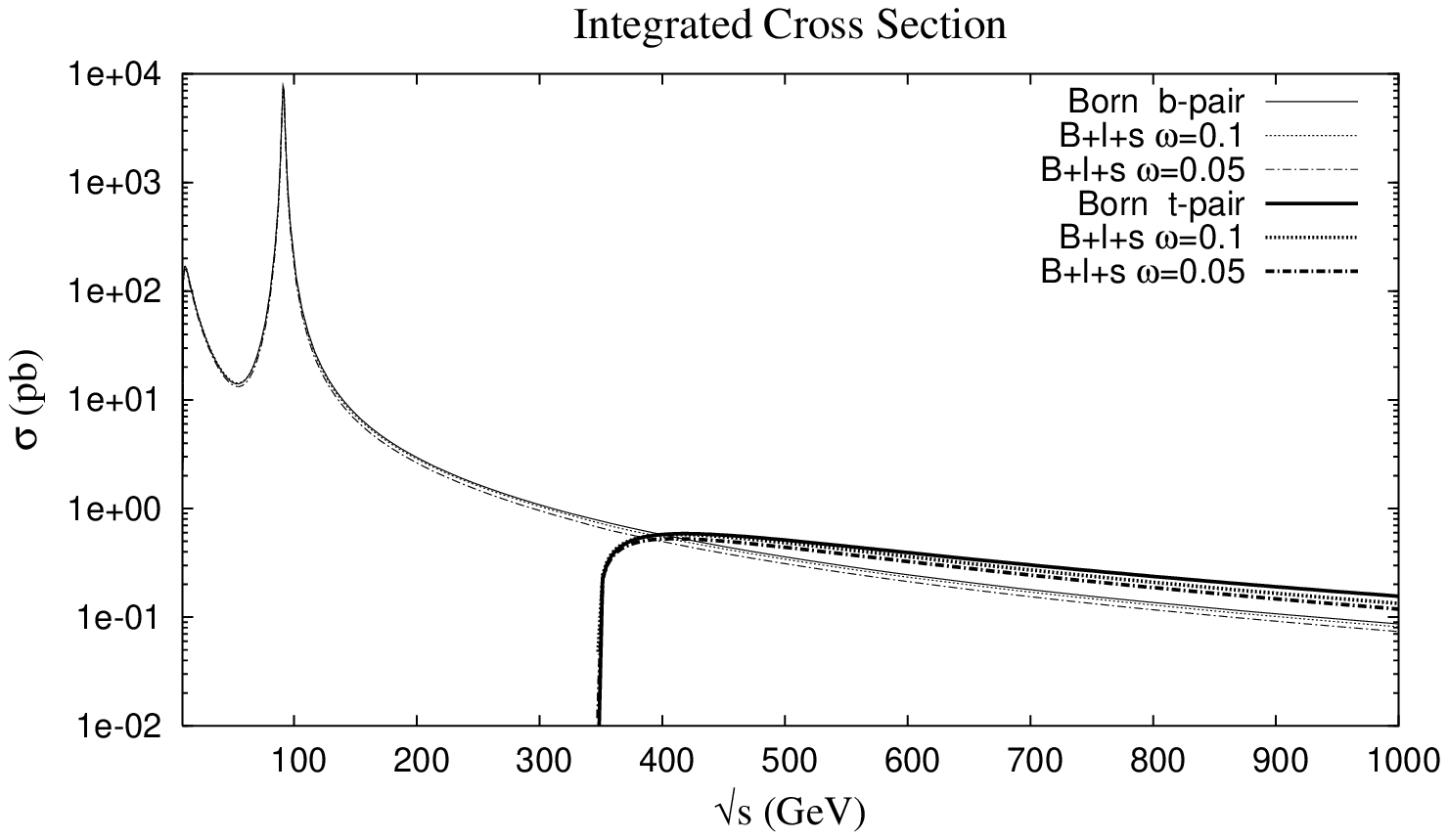}\\
\vspace{1em}
\includegraphics[scale=0.9]{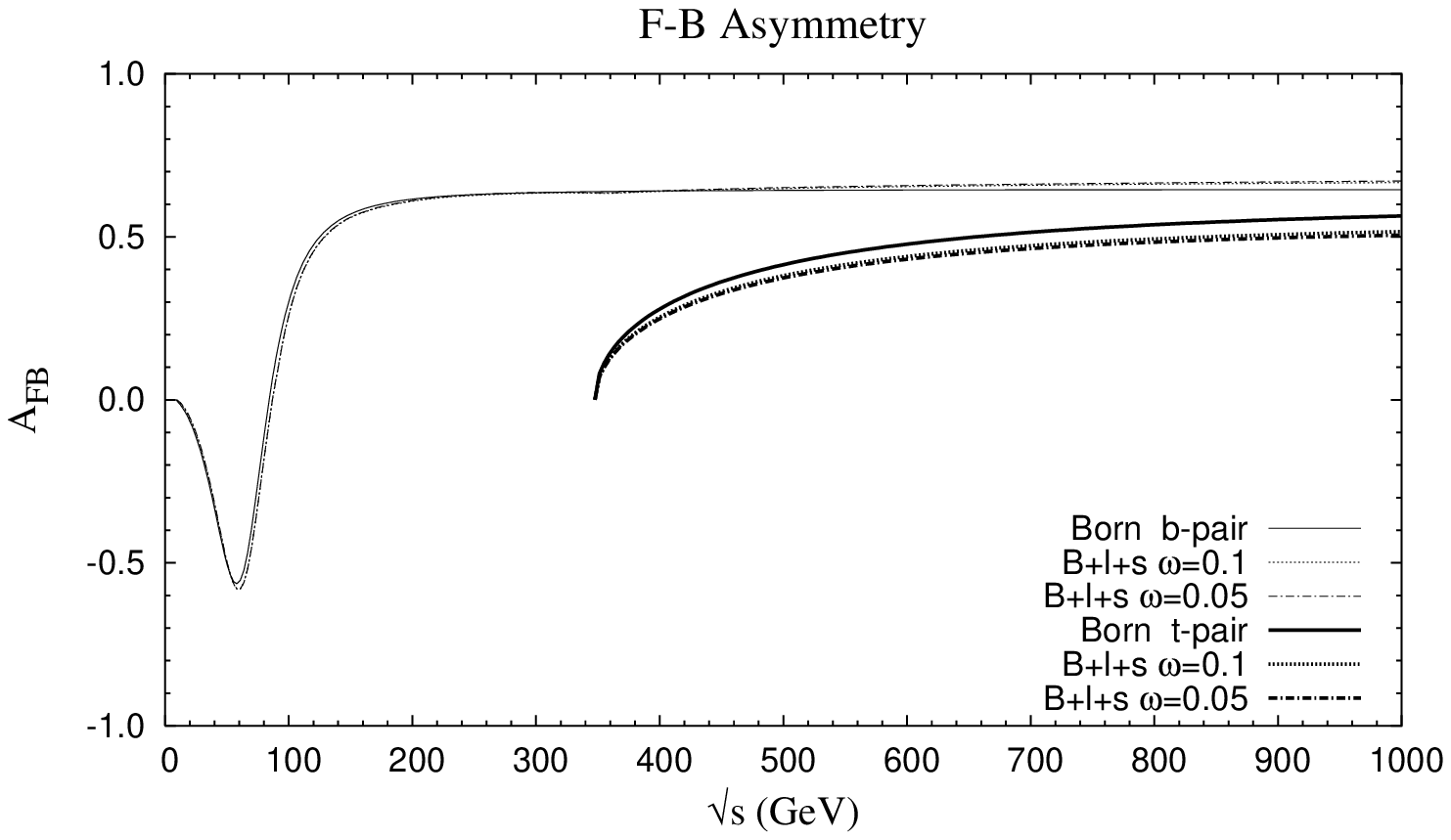}
\caption{Integrated cross section (above) and Forward-backward asymmetry (below) for $b\bar{b}$ (thin lines) and $t\bar{t}$ (thick ones) production at high energies. The corrected results (B+l+s) are computed with two alternatives soft-photon maximum energy fraction, $\omega$= 0.1 and 0.05.}
\label{ics-bt}
\efig

Two plots are presented in Fig. \ref{ics-bt}, trying to emphasize the
magnitude of the corrections with increasing energy for the $t\bar{t}$
and $b\bar{b}$ production. The figure on top shows the expected
integrated cross section, for the whole phase space. Continuous lines
represent the calculation at Born level while the others, dotted and
dash-dotted lines, account for the corrections with two different cuts
on the maximum energy available for the soft-photon, which is related
with the acollinearity of the two-fermions in the final state. The lower graph,
scans the expected forward-backward asymmetry, up to 1 TeV. Note the
phenomenological implications of the $Z$-boson threshold, inverting
the asymmetry for the $b$ quarks and also leading to the well-known
resonance on the integrated cross section. For the top, the threshold
energy for production is above the $Z$ mass leading to positive
asymmetries.

\subsection{Massive Bhabha scattering}

Besides the phenomenological implications of Bhabha scattering ($e^+
e^- \rightarrow e^+ e^-$) in accelerator physics, it has also been
considered here for `calibration' in automated computations. The fact
of comparing the massive case introduce most of the potential sources
for discrepancies between two different implementations.

\def\vertspace{-0.2ex}
\def\hahn{{\sl FA+FC}}
\begin{sidewaystable}[!th]
\caption{Numerical comparison of Bhabha scattering at $\sqrt{s}=500$
  GeV against the package \protect{\FA+\FC} and the massless electron case. Numbers for \protect\aitalc\ are shown in quadruple precision.}
\label{comparisonhahn}
\begin{displaymath}
{
\begin{array}{llll}
\hline 
\cos{\theta} \phantom{\Big|}
& {\left[ \frac{\mathrm{d} \sigma}{\mathrm{d} \cos{\theta}}
  \right]}_{\mathrm{Born}}\mathrm{/pb}
& {\left[ \frac{\mathrm{d} \sigma}{\mathrm{d} \cos{\theta}}
  \right]}_{\mathcal{O}(\alpha^3)=\textnormal{Born+QED+weak+soft}}\mathrm{/pb}
& \mathrm{Group} \\
\hline
\vspace{\vertspace}
  -0.9999 & 0.21482~70434~05632~80532~92137~22041~522\phantom{\cdot 10^0} & 0.14889~12189~28380~81306~77293~15050~269\phantom{\cdot 10^0}& \text{\aitalc}\\
  -0.9999 & 0.21482~70434~05633\phantom{\cdot 10^0} & 0.14889~12189~284\red{04}~\phantom{00000~00000~00000~000}\phantom{\cdot 10^0}& \hahn\\
\hline
\vspace{\vertspace}
  -0.9 & 0.21699~88288~10920~47030~98714~36154~378\phantom{\cdot 10^0} & 0.19344~50785~26862~70315~89749~62997~896 \phantom{\cdot 10^0}& \text{\aitalc}\\
\vspace{\vertspace}
  -0.9 & 0.21699~88288~10920\phantom{\cdot 10^0} & 0.19344~50785~26862~\phantom{00000~00000~00000~000}\phantom{\cdot 10^0}& \hahn\\
  -0.9 & 0.21699~88288~\red{41513}\phantom{00000~00000~00000~000}\phantom{\cdot 10^0} & 0.19344~50785~\red{62638}\phantom{\cdot 10^0}&{m_e=0} \\
\hline
\vspace{\vertspace}
  +0.0 & 0.59814~23072~50330~72189~85566~54984~067\phantom{\cdot 10^0} & 0.54667~71794~69423~03528~77433~99277~261 \phantom{\cdot 10^0}& \text{\aitalc}\\
\vspace{\vertspace}
  +0.0 & 0.59814~23072~5032\red{9}\phantom{\cdot 10^0} & 0.54667~71794~6942\red{2}~\phantom{00000~00000~00000~000}\phantom{\cdot 10^0}& \hahn\\
  +0.0 & 0.59814~23072~\red{88584}\phantom{\cdot 10^0} & 0.54667~71794~\red{99961}\phantom{\cdot 10^0}&{m_e=0} \\
\hline
\vspace{\vertspace}
  +0.9 & 0.18916~03223~32270~65918~47771~84443~698\cdot 10^3 & 0.17292~83490~66508~29307~47453~57027~839 \cdot 10^3& \text{\aitalc}\\
\vspace{\vertspace}
  +0.9 & 0.18916~03223~32271~\phantom{00000~00000~00000~000}\cdot 10^3 & 0.17292~83490~66508~\phantom{00000~00000~00000~000}\cdot   10^3& \hahn \\
  +0.9 & 0.18916~03223~3\red{1849}~\phantom{00000~00000~00000~000}\cdot 10^3 & 0.17292~83490~6\red{1347~4}\cdot 10^3 &{m_e=0} \\
\hline
\vspace{\vertspace}
  +0.9999 & 0.20842~90676~46390~74378~73935~63676~670 \cdot 10^9 & 0.19140~17861~11883~04292~09879~74774~038 \cdot 10^9& \text{\aitalc}\\
  +0.9999 & 0.20842~90676~464\red{36}~\phantom{00000~00000~00000~000}\cdot 10^9 & 0.19140~17861~11\red{979}~\phantom{00000~00000~00000~000}\cdot 10^9& \hahn \\
\hline
\end{array}
}
\end{displaymath}
\end{sidewaystable}

In Table \ref{comparisonhahn}, we present values for
differential cross sections at 500 GeV for different
$\cos{\theta}$. The expectations of the Born level and
$\mathcal{O}(\alpha)$ radiative corrected values are shown for the package
\aitalc \footnote{Quadruple precision is turned on here, setting 33 digits of precision.}, another totally independent automated
package composed out of \FA+\FC+\LT, and the same calculation droping out terms proportional to electron
mass. The results are impressive since the agreement between
both packages are far from the massless electron limit, and they
are limited by the roundoff of double-precision in the table. This
increases from $\sim$11 to 15 digits the previous agreement reported by
different groups in such a technical comparisons
\cite{Fleischer:2002nn,Hahn:2003ab,Andonov:2002jg}, performing
therefore a strong check of the correctness of the calculations.

\section{Conclusions}

We have focussed on the development and immediate applicability of
computer algebra and numerical tools to the scattering of two by two
fermions, resulting into a major initiative for fully automated
calculations, \aitalc, involving the feasibility of integrating the
work performed by others and to adapt it to some specific needs.

The direct application of this tool led to the calculation of
practically all $2 \rightarrow 2$ fermion processes of present
interest to $\mathcal{O}(\alpha)$. Still these predictions would
require further treatment to meet experimental observations, such as
kinematical cuts, strahlung effects or QCD corrections.

\section*{Acknowledgements}
We thank T. Hahn for producing a Bhabha \fortran\ code and for
communications when we performed numerical comparisons with this code.
We also thank M. Tentyukov for his hints while adapting \diana\ to our requirements.


\end{document}